\documentclass[12pt]{article}
\usepackage{amsmath,amssymb,epsfig}
\usepackage{graphicx}
\usepackage{cite}

\def\CPPP{\mathbb{C}\mathrm{P}^3}

\def\CP{\mathbb{C}\mathrm{P}^1}
\def\bF{\mathbf{F}}

\def \Tr{{\rm{Tr}}}
\def \xp{x^{+}}
\def \dxp2{dx^{+2}}
\def \xm{x^{-}}

\def \compsym{SU(2)\times SU(2)\times U(1) \times \mathbb{Z}_2}

\makeatletter

\renewcommand\section{\@startsection {section}{1}{\z@}%
                                   {-3.5ex \@plus -1ex \@minus -.2ex}%
                                   {2.3ex \@plus.2ex}%
                                   {\normalfont\large\bfseries}}

\renewcommand\subsection{\@startsection{subsection}{2}{\z@}%
                                     {-3.25ex\@plus -1ex \@minus -.2ex}%
                                     {1.5ex \@plus .2ex}%
                                     {\normalfont\normalsize\bfseries}}

\newcount\hour \newcount\minute
\hour=\time \divide \hour by 60
\minute=\time
\def\now{%
\ifnum \hour<13
  \ifnum \hour=0 \advance \hour by 12 \number\hour:\else \number\hour:\fi%
     \ifnum \minute<10 0\fi%
     \number\minute%
\ A.M.%
\else \advance \hour by -12 \number\hour:%
  \ifnum \minute<10 0\fi%
  \number\minute%
  \ P.M.%
\fi%
}

\makeatother

\begin{document}

\baselineskip=18pt  
\numberwithin{equation}{section}  
\allowdisplaybreaks  



%
%


\thispagestyle{empty}

\vspace*{-2cm}
\begin{flushright}
{\tt arXiv:0905.1954}\\
CALT-68-2731\\
IPMU-09-0061
\end{flushright}

\vspace*{1.7cm}
\begin{center}
 {\large {\bf Supersymmetric non-relativistic geometries in M-theory}}

 \vspace*{2cm}
 Hirosi Ooguri $^{*,\dagger}$
 and Chang-Soon Park $^*$\\
 \vspace*{1.0cm}
 $^*$
{\it California Institute of Technology 452-48, Pasadena, CA 91125, USA}\\
~~\\
 $^\dagger$ {\it Institute for the Physics and Mathematics of the Universe, \\
University of Tokyo, Kashiwa 277-8582, Japan}\\[1ex]
 \vspace*{0.8cm}
\end{center}
\vspace*{.5cm}

\noindent
We construct M-theory supergravity solutions with the non-relativistic Schr\"odinger symmetry starting from the warped $AdS_5$ metric with $\mathcal{N}=1$ supersymmetry.
We impose the condition that the lightlike direction is compact by making it a non-trivial $U(1)$ bundle over the compact space.
Sufficient conditions for such solutions are analyzed.
The solutions have two supercharges for generic values of parameters, but the number of supercharges increases to six in some special cases.
A Schr\"odinger geometry with $SU(2)\times SU(2) \times U(1)$ isometry is considered as a specific example.
We consider the Kaluza-Klein modes and show that the non-relativistic particle number is bounded above by the quantum numbers of the compact space.

\newpage
\setcounter{page}{1} 





\section{Introduction}
AdS/CFT correspondence has made a remarkable development in the past decade \cite{Maldacena:1997re,Witten:1998qj,Gubser:1998bc}.
The correspondence between $\mathcal{N}=4$ super Yang-Mills theory and multiple D3-branes is the first and most widely studied example.
On the other hand, the correspondence for multiple M2-branes has been quite mysterious until recently.
The situation changed when Bagger and Lambert\cite{Bagger:2007jr,Bagger:2006sk,Bagger:2007vi} discovered $\mathcal{N}=8$ Chern-Simons-matter theory(see also \cite{Gustavsson:2007vu}), by developing the idea of \cite{Schwarz:2004yj}.
However, it was difficult to increase the rank of the gauge group.
This is in some sense related to the fact that the maximally supersymmetric M2-brane solution does not have an adjustable parameter.
Later, Aharony et al. \cite{Aharony:2008ug} constructed $\mathcal{N}=6$ $U(N)\times U(N)$ Chern-Simons-matter theories that describe multiple M2-branes on the orbifold $\mathbb{C}^4/\mathbb{Z}_k$, where $k$ becomes the level of the Chern-Simons action in the field theory.
This orbifold provides us with an adjustable parameter, which enables us to treat weakly coupled field theories in some limit.

For multiple M2-branes in flat space, we can turn on an anti-self-dual four-form flux, which corresponds to adding a fermionic mass term to the field theory.
The four-form flux polarizes M2-branes into M5-branes\cite{Bena:2000zb,Bena:2004jw} and the discrete set of vacua of the theory has a one-to-one correspondence with the partition of $N$, the number of M2-branes\cite{Lin:2004nb}. For multiple M2-branes on the orbifold $\mathbb{C}^4/\mathbb{Z}_k$, we can consider a similar story.
A mass-deformed version of ABJM theory was considered in \cite{Hosomichi:2008jb} and its vacuum structure was identified in \cite{Gomis:2008vc}.
Especially, in the most symmetric vacuum, the system has $\compsym$ symmetry.
The mass term breaks the relativistic scaling symmetry.
However, there is a non-relativistic limit of this theory that has the Schr\"odinger symmetry\cite{Nakayama:2009cz,Lee:2009mm}.
Note that the Chern-Simons-matter theory is a good model to study the non-relativistic limit since gauge fields are not propagating.

Therefore, it is natural to seek for a supergravity solution that corresponds to the non-relativistic limit of the mass-deformed ABJM theory.
Assuming the classical analysis of the vacuum structure of the field theory is still applicable to the supergravity limit, the solutions will have $\compsym$ global symmetry and several additional $U(1)$ symmetries corresponding to the non-relativistic particle number symmetry, depending on which fields to retain in the non-relativistic limit\cite{Nakayama:2009cz}.
In the most supersymmetric case, it has 14 supercharges.
Although we were not able to find a supergravity solution with the same number of supersymmetries, we will present a class of supersymmetric solutions with the Schr\"odinger symmetry in two space dimensions in M-theory, and then consider a specific case with the same global bosonic symmetry of the non-relativistic limit of the mass-deformed ABJM theory.

A geometry with the Schr\"odinger symmetry was found in \cite{Son:2008ye,Balasubramanian:2008dm}\footnote{See \cite{Duval:1990hj} for an earlier discussion, whose relation is explained in \cite{Duval:2008jg}.}.
In this case, the $AdS$ symmetry is explicitly broken to the Schr\"odinger symmetry due to the term $-\frac{\dxp2}{r^4}$ in the metric, where $\xp$ is one of the two lightlike coordinates.
Soon after, the geometry was embedded in string theory \cite{Adams:2008wt,Maldacena:2008wh,Herzog:2008wg}.
The supergravity solutions with the Schr\"odinger symmetry does not have supersymmetry mainly due to the term $-\frac{\dxp2}{r^4}$ in the metric and the lightlike three form flux $H_3$ that supports it.
Supersymmetry can be recovered if the coefficient of $\frac{\dxp2}{r^4}$ depends on the compact space\cite{Hartnoll:2008rs}.
However, in their case, the coefficient is necessarily negative in some region of the compact space and the stability of the spacetime is not guaranteed.
Recently, this problem was remedied and supersymmetric solutions were obtained with negative coefficient of $\frac{\dxp2}{r^4}$ by turning on some lightlike fluxes, which can be related either to a Killing vector that leaves some Killing spinors invariant\cite{Bobev:2009mw}, or to the properties of the Calabi-Yau structure\cite{Donos:2009xc}.
Also, it is possible to explicitly break the $AdS$ symmetry by adding a term $d\xp C$ to the metric where $C$ is a one-form that does not depend on the worldvolume coordinates\cite{Donos:2009en,Donos:2009xc}.
There are also proposals where the breaking occurs due to the fact that the lightlike direction is compact without explicitly adding a term to the $AdS$ metric\cite{Goldberger:2008vg,Barbon:2008bg}.

In this paper, we will explore supergravity solutions having the Schr\"odinger symmetry in M-theory.
Since a non-relativistic field theory has a discrete particle number, we expect the $U(1)$ direction corresponding to the particle number is compact.
Instead of imposing this as an additional assumption, we make the compact lightlike direction a non-trivial $U(1)$ bundle over the compact space.
Then the compactness is required without further assumption.
We begin with the $\mathcal{N}=1$ warped $AdS_5$ solutions in M-theory given in  \cite{Gauntlett:2004zh}, and modify the geometry to obtain the Schr\"odinger symmetry.
Initially the $AdS_5$ solution has eight supercharges and they reduce to two after the modification in general.
However, there is a special case when there remain six supercharges, which is the same number as in the DLCQ of $AdS$.
After general remarks, we specialize to a specific example with $SU(2)\times SU(2)\times U(1)$ isometry.
We consider the Kaluza-Klein spectrum of the theory, and show that the non-trivial $U(1)$ bundle structure of the lightlike compact direction sets an upper bound for the non-relativistic particle number for given quantum numbers of the compact space.
The initial motivation to consider a Schr\"odinger invariant geometry with $SU(2)\times SU(2)\times U(1)$ was to find a candidate theory for the dual of the non-relativistic mass-deformed ABJM theory.
In line with this, we also provide a non-supersymmetric solution with the same global symmetry briefly at the end.

\section{General consideration}
In this section, we will deform the supergravity solutions given in \cite{Gauntlett:2004zh} in such a way that the resulting solutions have the Schr\"odinger symmetry.

\subsection{Warped $AdS_5$ solutions in M-theory}
Before dealing with non-relativistic solutions, let us describe the general $\mathcal{N}=1$ supersymmetric solutions of the supergravity limit of M-theory consisting of a warped product of $AdS_5$ and a six-dimensional space considered in \cite{Gauntlett:2004zh}.
The metric is of the form
\begin{equation}
ds^2= e^{2\lambda} \left[ ds^2_{AdS_5} + ds^2_{M_6}\right]
\end{equation}
and the four-form flux lies along the compact six dimensions.
The overall coefficient $e^{\lambda}$ is a warping factor that depends on $M_6$.
The authors of \cite{Gauntlett:2004zh} obtained the most general condition for $\mathcal{N}=1$ supersymmetry, and then specialized to a special case where the six-dimensional manifold $M_6$ is a complex manifold with a Hermitian metric.
In this case, the supersymmetry condition becomes significantly simplified and they can obtain many explicit solutions.
Let us describe the manifold $M_6$ first.
The metric of $M_6$ is given by
\begin{equation}
ds^2_{M_6}=e^{-6\lambda(y)} \left[ \hat{g}_{ij}(x,y) dx^i dx^j + \sec^2 \zeta(y) dy^2 \right] + \frac 1 9 \cos^2 \zeta(y) (d\psi+\hat P)^2\;.
\end{equation}
There is a four dimensional K\"ahler manifold $M_4$, whose metric is $\hat{g}_{ij} dx^i dx^j$.
The complex structure of the metric is independent of $y$ and $\psi$.
$\frac{\partial}{\partial \psi}$ is a Killing vector of $M_6$ and the $y$ dependence of the metric warps the spacetime.
$\hat P$ is the canonical Ricci-form connection defined by the K\"ahler metric $\hat g$.
That is, the Ricci form $R=d\hat P$.
$\hat P$ is independent of $y$ and $\psi$.
$\zeta$ is a function of $y$ which is implicitly defined by
\begin{equation}\label{E:ylambdazeta}
2y=e^{3\lambda} \sin \zeta\;.
\end{equation}
We fix the $AdS_5$ radius to be 1.
The four-form field strength is given by
\begin{equation}\label{E:genAdSF4}
\begin{split}
F_4^{(0)}&=-(\partial_y e^{-6\lambda}) \hat V_4 + \frac 1 3 dy\wedge (d\psi+\hat P)\wedge \hat L\\
\hat  L&= \frac 1 3 \cos^2\zeta \hat{*}_4 d \hat P - 4 e^{-6\lambda} \hat J\;,
\end{split}
\end{equation}
where $\hat V_4$ is the volume form and $\hat J$ is the K\"ahler form of $M_4$.
In addition to these, we have two more constraints:
\begin{equation}
\begin{split}
\partial_y \hat J &= -\frac 2 3 y d \hat P\\
\partial_y \log \sqrt{\hat g} &= -3 y^{-1} \tan^2 \zeta -2 \partial_y \log\cos\zeta\;.
\end{split}
\end{equation}
Given these conditions, the Bianchi identity and the equations of motion for $F_4^{(0)}$, and the Einstein equations are all satisfied.

\subsection{Deformation to solutions with Schr\"odinger symmetry}\label{SS:DeftoSch}
Let us first write the $AdS_5$ metric in a form that will be suitable for later analysis:
\begin{equation}
ds^2_{AdS_5} = \frac{-2 d\xp d\xm + dx_1^2 + dx_2^2 +dr^2}{r^2}\;.
\end{equation}
The DLCQ of $AdS_5$ makes the $\xm$ direction compact.
The modification we do here is to make $x^-$ a coordinate for a $U(1)$ bundle over the compact space.
In the case when the $U(1)$ bundle is non-trivial, the lightlike direction is necessarily compact and breaks $AdS_5$ symmetry down to the Schr\"odinger symmetry\footnote{There was a paper\cite{Colgain:2009wm} that also considers modification of the warped $AdS_5$ solutions of \cite{Gauntlett:2004zh}.
They added $dx^+ C$ component to the metric, where $C$ is a globally defined one-form on the compact space, which means the $U(1)$ bundle corresponding to the the $x^-$ direction is trivial.}.
Let us call the geometry $Sch_5$.

Note that making the lightlike direction compact makes it subtle to deal with the system in the supergravity approximation.
The situation gets better if we add large momenta along the compact lightlike direction\cite{Maldacena:2008wh}.
This will involve making a black hole solution that asymptotes to the geometry that we give below.
We will not consider such a finite temperature/finite density solution here, but we note that the compact lightlike direction changes the causal structure of the spacetime drastically.
In particular, any two points in the geometry can be joined by a timelike or lightlike curve:
Suppose we want to connect some point $P=(\xp, \xm, x^i, r)$ to $Q=(0,0,0,0)$ using a timelike curve when $\xp<0$.
Due to the periodic identification, we can equally start at $P=(\xp, \xm- N \Delta \xm , x^i, r)$ for some large $N$ where $\Delta \xm$ is the period of the $\xm$ direction.
For large enough $N$, there is indeed a timelike curve connecting the points $P$ and $Q$.
This is a property that is expected for the dual theory of a non-relativistic system.

Note that we can also add a term proportional to $\frac {\dxp2}{r^4}$, which does not break the Schr\"odinger symmetry.
The coefficient depends on the compact space.
Such a possibility was explored previously in \cite{Hartnoll:2008rs}.
Specifically, we consider the following metric:
\begin{equation}\label{E:Sch5Met}
\begin{split}
ds^2&= e^{2\lambda} \left[ ds^2_{Sch_5} + ds^2_{M_6}\right]\\
ds^2_{Sch_5}&=-f(y)\frac{\dxp2}{r^4}+\frac{-2 d\xp (d\xm+ A) +dx_1^2+dx_2^2+dr^2}{r^2}\\
ds^2_{M_6}&=e^{-6\lambda(y)} \left[ ds^2_{M_4} + \sec^2 \zeta(y) dy^2 \right] + \frac 1 9 \cos^2 \zeta(y) (d\psi+\hat P)^2\\
ds^2_{M_4}&=\hat g_{ij}(x,y) dx^i dx^j\;.
\end{split}
\end{equation}
$A$ is a gauge field on $M_4$ and $f(y)$ is some function that depends only on $y$.
We need to determine these two quantities.
To support this geometry, we turn on the four-form field strength along the lightlike direction:
\begin{equation}
F_4=F_4^{(0)}+ \frac 1 {r^3} s(y) d\xp \wedge dr \wedge dA - \frac 1 {2r^2} s'(y) d\xp \wedge dy \wedge dA\;.
\end{equation}
We demand that $A$ depends only on $x^i$, and not on $y$: otherwise, the second term includes a part proportional to $d\xp\wedge dr\wedge dy\wedge \partial_y(dA)$, and then it is impossible to satisfy the equations of motion for $F_4$.
$F_4^{(0)}$ is the original four-form field strength of the warped $AdS_5$ solution, and $s(y)$ is some function to be determined.
By construction, $dF_4=0$.
Just as in the original warped $AdS_5$ solution, we also require $F_4\wedge F_4=0$.
This requires
\begin{equation}
\hat L \wedge dA=0\;.
\end{equation}
Let us consider the equations of motion for $F_4$ first.
The dual seven-form $F_7$ is given by
\begin{equation}
\begin{split}
F_7=*_{11} F_4 = &F^{(0)}_7 + e^{6\lambda} \frac{1}{r^5}d\xp \wedge dx_1 \wedge  dx_2 \wedge  dr \wedge  A \wedge \left[ 2 \lambda'(y) dy\wedge (d\psi+\hat P) + \hat{*}_4 \hat L\right]\\
&+\frac {s(y)}{3 r^4} d\xp \wedge  dx_1 \wedge  dx_2  \wedge dy  \wedge (d\psi+ \hat P)  \wedge \hat{*}_4 dA\\
&+\frac {s'(y)}{6r^5} e^{6\lambda}\cos^2 \zeta d\xp \wedge  dx_1  \wedge dx_2  \wedge dr \wedge  (d\psi+ \hat P)  \wedge \hat{*}_4 dA\;,
\end{split}
\end{equation}
where $F^{(0)}_7$ is the seven-form field strength of the corresponding warped $AdS_5$ solution.
Since we only consider the case when $F_4\wedge F_4=0$, the equation of motion of $F_4$ is satisfied when $d F_7=0$.
This is satisfied provided
\begin{equation}\label{E:ExtofSol}
\begin{split}
d A&=\pm\hat{*}_4 d A\\
d \hat P \wedge \hat{*}_4 d A&=0\\
dA\wedge \hat{*}_4 \hat L&=0
\end{split}
\end{equation}
as well as
\begin{equation}\label{E:EOMDF40}
\pm 12 e^{6\lambda} \lambda' + 8 s(y) + \partial_y (s'(y) e^{6\lambda}\cos^2 \zeta) =0\;.
\end{equation}
The last equation is satisfied when
\begin{equation}
\begin{split}
s(y)&=-2y\qquad \text{if $dA$ is self-dual}\\
s(y)&=2y\qquad \text{if $dA$ is anti-self-dual}\;.
\end{split}
\end{equation}
due to the relation \eqref{E:ylambdazeta}.
In the cases we are interested, $y$ takes values between two roots of $\cos\zeta=0$.
Since \eqref{E:EOMDF40} is a second order differential equation and the coefficient of $s''(y)$ vanishes when $\cos\zeta=0$, the other solution necessarily blows up when $\cos\zeta=0$.
Therefore, $s(y)=2y$ is the regular solution we want.
The third equation implies $dA\wedge \hat J=0$.
We will see presently that the Einstein equations are also satisfied by choosing the coefficient $f(y)$ of $\frac {\dxp2}{r^4}$ appropriately.
However, it is possible that the coefficient can take both positive and negative values over the compact space and, in the example that we consider in the next section, indeed this is the case.
This is analogous to the situation considered in \cite{Hartnoll:2008rs}, where the coefficient of $\frac {1}{r^4}\dxp2$ is a harmonic function, which implies that it is necessarily negative in some region of the compact space.
They show that there is an instability of a field with sufficiently large particle number due to the unboundedness of the Hamiltonian $H$(the conjugate momentum to $\xp$).
Supersymmetry cannot guarantee $H$ is positive since there is no dynamical supercharge.
We expect a similar instability in our geometry unless $f(y)$ vanishes.
However, as we will see in section \ref{E:genSUSY}, when $f(y)=0$, there are two dynamical supercharges and the Hamiltonian $H$ is bounded by the condition $\{Q, Q^{\dagger}\}=H$ for dynamical supercharges $Q$ and $Q^{\dagger}$.

To sum up, if there is a harmonic (anti)self-dual two-form $d A$ that satisfies
\begin{equation}\label{E:EOMcond}
d \hat P\wedge dA=0\;,\qquad \hat J \wedge dA=0\;,
\end{equation}
then we can construct a supergravity solution with the Schr\"odinger symmetry as described above
\footnote{$dA$ represents a non-trivial element of the second cohomology class $H^2(M_4)$.
For this to be a non-trivial element of $H^2(M_6)$, we need to assume a global structure of the six-dimensional complex manifold $M_6$.
In the examples of \cite{Gauntlett:2004zh}, $M_6$ is taken to be a $\CP$ bundle over the K\"ahler base $M_4$.
Then the Gysin sequence $0\rightarrow H^2(M_4)\rightarrow H^2(M_6)$ implies $dA$ is also a non-trivial element of $H^2(M_6)$ as long as the orientability condition is satisfied.}.
Note that $A$ is a one-form on $M_4$ and does not depend on $y$.
Since $\partial_y \hat J = -\frac 2 3 y d \hat P$ and $\hat P$ is independent of $y$, if \eqref{E:EOMcond} is satisfied at one $y$, it is automatically satisfied for all $y$.

One case where a solution is easily found is when the manifold $M_4$ is K\"ahler-Einstein and $y$ and $\psi$ give a $\CP$ bundle over $M_4$.
The isometry of $\CP$ is broken to $U(1)$ by the warping factor that depends on $y$.
In this case, $d\hat P$, the Ricci form, is proportional to $\hat J$.
Since $d\hat P$ is $y$-independent, $\hat J$ factorizes into a $y$-dependent function and a $y$-independent form.
Hence, given a harmonic (anti)self-dual two-form $d A$ with $\hat J \wedge d A=0$, we can construct a Schr\"odinger solution.
To do that, the dimension of the second cohomology class has to be greater than 1, which means we cannot construct our solution on $\CPPP$.
However, there are cases when the dimension of the second cohomology class is greater than 1, and we will consider such an example where the manifold $M_4$ is $S^2\times S^2$.

Given the above requirement, the Einstein equations are satisfied by choosing a suitable $f(y)$.
Let us first introduce the following vielbeins:
\begin{equation}
\begin{split}
&E^{0}=e^{\lambda}\left(\frac{1+f(y)}{2} \frac 1 {r^2} d\xp+d\xm+A\right)\\
&E^{1}=e^{\lambda} \frac 1 r dx_1\;,\qquad E^{2}=e^{\lambda} \frac 1 r dx_2\\
&E^{3}=e^{\lambda}\left(\frac{1-f(y)}{2} \frac 1 {r^2} d\xp-(d\xm+A)\right)\\
&E^{4}=e^{\lambda}\frac{dr}{r}\\
&E^{y}=e^{-2\lambda}\sec\zeta dy\;,\qquad E^{\psi}=\frac 1 3 e^{\lambda}\cos\zeta (d\psi+\hat P)\\
&E^{i}=e^{-2\lambda} \hat e^i \;,\qquad i=1,2,3,4\;,
\end{split}
\end{equation}
where $\hat e^i$ are vielbeins for the metric $ds^2_{M_4}$ in \eqref{E:Sch5Met}.
Knowing that the original warped $AdS_5$ solution satisfies the Einstein equations of motion, all we need to check is the change of the component $G_{03}=\kappa^2_{11} T_{03}$ of the Einstein equation.
This will be satisfied if
\begin{equation}
-f(y)+ y f'(y)-\frac 1 {12} e^{6\lambda} \cos^2\zeta f''(y)=0\;.
\end{equation}
There are two linearly independent solutions and one obvious solution is $f(y)=\beta y$ for an arbitrary constant $\beta$.
In the case when $y$ and $\psi$ combine to give topologically a two-sphere $S^2$, $\cos\zeta=0$ at the two poles of the sphere, and we take the solution $f(y)=\beta y$ as the smooth solution.
The other solution diverges when $\cos\zeta=0$.

\subsection{Supersymmetry}\label{E:genSUSY}

The Killing spinor equation is given by
\begin{equation}\label{E:KSeqinM}
\delta \Psi_A = D_A \epsilon =\nabla_A \epsilon + \frac 1 {12} \left(\Gamma_A \bF^{(4)} - 3 \bF^{(4)}_A\right) \epsilon=\partial_A \epsilon + \frac 1 4 \omega_{ABC}\Gamma^{BC} + \frac 1 {12} \left(\Gamma_A \bF^{(4)} - 3 \bF^{(4)}_A\right) \epsilon\;,
\end{equation}
where $\epsilon$ is a Killing spinor and
\begin{equation}
\begin{split}
\bF^{(4)}&=\frac 1 {4!} F_{ABCD}\Gamma^{ABCD}\\
\bF^{(4)}_A &= \frac 1 2 \left[ \Gamma_A , \bF^{(4)}\right]\;.
\end{split}
\end{equation}
We use $A,B,\cdots$ for vielbein indices and $M,N,\cdots$ for coordinate indices of eleven dimensions.
Our strategy is to divide the operator $D_A$ into two: one is independent of $\beta$ and $A$, while the other is not.
Then, given a Killing spinor $\epsilon$ of the corresponding $AdS$ solution, we impose the condition that $\epsilon$ is annihilated by $\beta,A$-dependent part.
Let us denote by $\Delta \partial_A$ the change of the derivative $\partial_A$ due to the presence of $\beta$ and $A$, and similarly denote by $\Delta \omega_{A}$ the change of the connection $\omega_{ABC} \Gamma^{BC}$.
If we define a matrix $\Lambda^{A}_{\,\,\,\,M}$ by $E^A=\Lambda^{A}_{\,\,\,\,M} dx^M$, ${(\Lambda^{T})^{-1}}_A^{\,\,\,\,M}\partial_M=\partial_A$.
Then it is easy to see that the only components that depend on $\beta$ are ${(\Lambda^{T})^{-1}}_0^{\,\,\,\,-}$ and ${(\Lambda^{T})^{-1}}_3^{\,\,\,\,-}$, and those that depend on $A$ are ${(\Lambda^{T})^{-1}}_i^{\,\,\,\,-}$.
Therefore, we keep Killing spinors of the $AdS$ solution when it is independent of $\xm$. 
We will see later that the Killing spinors consistent with the compactification of $\xm$ are all independent of $\xm$.
Hence it does not give a new condition.

Next, let us consider the change of the connection $\Delta \omega_{A}$.
They are given by
\begin{equation}\label{E:chinomega}
\begin{split}
&\Delta\omega_1=\Delta\omega_2=\Delta \omega_4=\Delta \omega_y=\Delta \omega_{\psi}=0\\
&\Delta \omega_0 = \Delta \omega_3=\beta e^{2\lambda}\left(-\sin\zeta \Gamma^4 +\cos\zeta \Gamma^{y}\right)\Gamma^+ + e^{5\lambda} \bF^{(2)}\\
&\Delta \omega_i = - e^{5\lambda} F_{ij} \Gamma^{j} \Gamma^{+}\;.
\end{split}
\end{equation}
Here $\Gamma^+=\Gamma^0+\Gamma^3$ and $\bF^{(2)}$ is a product of gamma matrices $\frac 1 2 F_{ij} \Gamma^{ij}$ where $F=dA$ and $\frac 1 2 F_{ij}\hat{e}^i\hat{e}^j=F$.
The change in the four-form field strength is
\begin{equation}\label{E:chinF4}
\Delta \bF^{(4)} = e^{5\lambda} \left(-\sin\zeta\Gamma^4+\cos\zeta \Gamma^y \right) \Gamma^{+} \bF^{(2)}\;.
\end{equation}
The condition that the differential operators $D_0$ and $D_3$ still annihilate a Killing spinor $\epsilon$ of the $AdS$ solution imposes
\begin{equation}\label{E:genericSUSYcons}
\beta \, \Gamma^+ \epsilon=0;\;,\qquad \bF^{(2)}\epsilon=0\;.
\end{equation}
The second equation is satisfied if, for example, the manifold $M_4$ is K\"ahler-Einstein and the two-form field strength $F$ is a $(1,1)$-form on $M_4$.
To see this, let us decompose gamma matrices and spinors into $AdS_5$ and $M_6$ parts(note that we are looking for a Killing spinor of the original $AdS_5$ geometry which survives after we change the metric to the $Sch_5$ geometry).
First, we decompose the eleven dimensional gamma matrices as
\begin{equation}
\begin{split}
\Gamma^a&=\gamma^a \otimes \tau_7\\
\Gamma^m&=1\otimes \tau^m\;,
\end{split}
\end{equation}
where $a=0,\cdots, 4$ and $m=1,\cdots, 6$ are orthonormal indices for $AdS_5$ and $M_6$, respectively, and $\tau_7=\tau_1\cdots \tau_6$.
They satisfy
\begin{equation}
\begin{split}
\{\gamma^a , \gamma^b \} &= -2 \eta^{ab}\\
\{\tau^m, \tau^n\} &= 2\delta^{mn}\;,
\end{split}
\end{equation}
where $\eta^{ab}=\mathrm{diag}(-1,1,1,1,1)$.
Note $\tau_7^2=-1$.

The Killing spinor $\epsilon$ is decomposed as $\psi(x)\otimes e^{\frac{\lambda}2}\xi(y)$ for $x\in AdS_5$ and $y\in M_6$.
$\psi$ satisfies the Killing spinor equation for $AdS_5$:
\begin{equation}
\partial_a \psi - \frac 1 4 \omega_{abc}\gamma^{bc} \psi = \frac 1 2 i \gamma_a \psi\;.
\end{equation}
There are two types of Killing spinors of $AdS_5$.
They are given as(see, for example, \cite{Lu:1998nu})
\begin{equation}\label{E:KSinAdS}
\psi^+=r^{-\frac 1 2} \psi_0^{+}\;,\qquad \psi^-=(r^{\frac 1 2}+i r^{-\frac 1 2} x^{\mu} \gamma_{\mu})\psi_0^{-}\;,
\end{equation}
where $-i \gamma^r \psi_0^{\pm} = \pm \psi_0^{\pm}$.
$\psi^+$ generates a Poincar\'e supersymmetry and $\psi^-$ a superconformal one.
In our case, the lightlike direction $\xm$ is compactified.
$\psi^+$ depends only on $r$, so $\psi^+$ survives compactification.
$\psi^-$ is position dependent, and to be periodic in $\xm$, it should not have $\xm$ dependence.
This is the same as requiring that $\gamma^+\psi^-_0=0$.
Hence half of the superconformal supersymmetries survive compactification.

The Killing spinor equation $D_a \epsilon=0$ implies that $\xi$ has to satisfy
\begin{equation}\label{E:Daepsilon}
\left(\tau^m \nabla_m \lambda + \frac 1 {6} e^{-3\lambda}\bF_{0}^{(4)}- i \tau_7\right)\xi=0\;.
\end{equation}
where $\bF_{0}^{(4)}$ is a gamma matrix expression using $\tau^m$ constructed from the four-form field strength \eqref{E:genAdSF4}.
Let us multiply the above equation by $\bF^{(2)}$, where now $\bF^{(2)}$ is made up of $\tau^m$ matrices:
\begin{equation}\label{E:bF2mul}
\bF^{(2)}\left(\tau^m \nabla_m \lambda + \frac 1 {6} e^{-3\lambda}\bF_{0}^{(4)}- i \tau_7\right)\xi=0\;.
\end{equation}
From \eqref{E:ExtofSol}, we obtain $\hat L \wedge F = \hat L \wedge \hat{*}_4 F=0$.
This implies $\{\bF^{(2)}, \mathbf{\hat{L}}\}=0$ since $\{\tau^{mn}, \tau^{pq}\}=2\gamma^{mnpq}-4\delta_{mn}^{pq}$.
Also, since we assume $M_4$ is K\"ahler-Einstein, $[\bF^{(2)}, \mathbf{\hat{L}}]$ is proportional to $F_{ij} \hat{J}^{j}_{\,\,\,k} \Gamma^{ik}$, which vanishes if $F$ is a (1,1)-form.
Now, we can simplify the expression \eqref{E:bF2mul} in the form $\mathbf Q \bF^{(2)} \xi=0$ where $\mathbf Q$ is some linear combination of gamma matrices.
By examining the explicit expression, we see that $\mathbf Q$ has determinant $(1-4y^2(\lambda')^2)^4$, which does not vanish.
Therefore we conclude $\bF^{(2)} \xi =0$.
The remaining constraints come from examining $D_i \xi=0$ for $i=\theta_1,\phi_1,\theta_2,\phi_2$.
In the case $\bF^{(2)} \xi =0$, they impose an additional condition
\begin{equation}\label{E:beta0SUSY}
\Gamma^{+}\left(1+ \sin \zeta \Gamma^{4} - \cos \zeta \Gamma^{y}\right) \epsilon = 0\;.
\end{equation}
This is satisfied if $\Gamma^+ \epsilon=0$.

In conclusion, a Killing spinor of the $AdS$ solution survive if it satisfies $\Gamma^+ \epsilon=0$.
Therefore, at each point, a Killing spinor has to lie in some four dimensional space.
This does not necessarily mean that there are four Killing spinors, since higher order integrability condition may not be satisfied.
In fact, a superconformal supercharge cannot satisfy $\Gamma^+ \epsilon=0$.
To see this, note that a superconformal supercharge is represented in the Poincar\'e coordinates as in the second expression in \eqref{E:KSinAdS}.
$\Gamma^+ \epsilon=0$ translates into $\gamma^+\psi^-=0$, which is written as
\begin{equation}
\gamma^+ \left[r^{\frac 1 2}+i r^{-\frac 1 2} (x^i \gamma^i - x^+ \gamma^- - x^- \gamma^+)\right]\psi_0^{-}=0\;.
\end{equation}
At $x=0$, this implies $\gamma^+\psi_0^-=0$.
Now, we move $\gamma^+$ to the right.
Then, since $\{\gamma^+, \gamma^-\}=2$, we end up getting $\psi_0^-=0$, which means the only solution to this equation is the trivial one.
Hence no superconformal supersymmetries survive, which means there remain only two Poincar\'e supercharges that are annihilated by $\gamma^+$.

When $\beta=0$, there can be more supercharges since the first equation of \eqref{E:genericSUSYcons} is trivial and all we require is \eqref{E:beta0SUSY} as well as $\bF^{(2)}\epsilon=0$.
We have already considered the case when $\Gamma^{+}\epsilon=0$.
Another possibility is that $\epsilon$ is annihilated by the second factor.
Under the decomposition, this can be rewritten as
\begin{equation}\label{E:rank16eqdecomp}
\left(1\pm i \sin\zeta \tau_7 -\cos \zeta \tau_y\right)\xi=0\;,
\end{equation}
depending on $-i \gamma^r \psi^{\pm} = \pm \psi^{\pm}$.

Now, we will prove that \eqref{E:Daepsilon} implies \eqref{E:rank16eqdecomp} with plus sign in the second term if $M_4$ is K\"ahler-Einstein and $F$ is anti-self-dual.
$\bF^{(2)}\bF^{(2)}\epsilon=0$ implies $\tau^{3456} \epsilon=-\epsilon$ if $F$ is anti-self-dual.
The indices for $M_6$ are such that $\{y,\psi,\theta_1,\phi_1,\theta_2,\phi_2\}\leftrightarrow \{1,2,3,4,5,6\}$.
For a K\"ahler-Einstein manifold $M_4$, $\hat L$ is given by\cite{Gauntlett:2004zh}
\begin{equation}\label{E:hatLexp}
\hat L=\left( \frac{\cos^2\zeta (1+ 6y \lambda')}{e^{6\lambda}-4y^2} - 4e^{-6\lambda}\right)\hat J\;.
\end{equation}
Define $\mathbf{\hat J}=\frac 1 2 e^{6\lambda} \hat{J}_{ij}\tau^{ij}$ where $\hat J=\frac 1 2 \hat{J}_{ij} \hat{e}^i\hat{e}^j$, and define $\mathbf{\hat L}$ similarly.
Since $\hat J$ is self-dual,
\begin{equation}\label{E:hatJexp}
\mathbf{\hat J}\mathbf{\hat J}= \frac 1 2 \{\mathbf{\hat J},\mathbf{\hat J}\}=e^{12\lambda}(\tau^{3456}-1)\;,
\end{equation}
We can rewrite \eqref{E:Daepsilon} as
\begin{equation}
\left(e^{3\lambda}\lambda' \cos\zeta \tau_1 + e^{3\lambda} \lambda' \tau_{3456}+\frac 1 6 \tau_{12}\mathbf{\hat{L}}-i\tau_7\right)\xi=0\;.
\end{equation}
By multiplying by $\tau_7 \left(e^{3\lambda}\lambda' \cos\zeta \tau_1 - e^{3\lambda} \lambda' \tau_{3456}+\frac 1 6 \tau_{12}\mathbf{\hat{L}}+i\tau_7\right)$ to the left, up to an overall factor, we obtain \eqref{E:rank16eqdecomp} with plus sign in the second term if we use \eqref{E:hatLexp}, \eqref{E:hatJexp} and $\tau^{3456} \epsilon=-\epsilon$.
That implies that the corresponding Killing spinor in the $AdS_5$ part is a Poincar\'e supercharge.
Therefore, when $\beta=0$, we have four Poincar\'e supercharges.

However, there should be additional supercharges that we might have overlooked when we analyze \eqref{E:beta0SUSY}.
Indeed, if we keep all four Poincar\'e supercharges of the $AdS$ solution, there are two kinematical supercharges and two dynamical ones\footnote{For a related discussion about Schr\"odinger superalgebra, see \cite{Leblanc:1992wu,Duval:1993hs,Sakaguchi:2008rx,Sakaguchi:2008ku}.}.
In this case, the commutator of the special conformal generator $C$ and a dynamical supercharge $Q$ produces a superconformal supercharge $S$: $[C,Q]\sim S$.
Therefore, there has to be a way to obtain a superconformal supersymmetry.
To see how it comes about, let us look at the expression for a superconformal supersymmetry in $AdS_5$ space
\begin{equation}
\psi^-=(r^{\frac 1 2}+i r^{-\frac 1 2} x^{\mu} \gamma_{\mu})\psi_0^{-}=\left[r^{\frac 1 2}+i r^{-\frac 1 2} (x^i \gamma^i - x^+ \gamma^- - x^- \gamma^+)\right]\psi_0^{-}
\end{equation}
with $-i \gamma^r \psi^-_0 = -\psi^-_0$.
Since $\xm$ is compactified, we demand $\gamma^+ \psi^-_0=0$.
If we set $\epsilon=\psi^- \otimes e^{\frac{\lambda}{2}} \xi(y)$ with $\psi^-$ as just given, \eqref{E:beta0SUSY} becomes
\begin{equation}
\left(1+ i \sin\zeta \tau_7 -\cos \zeta \tau_y\right)\xi=0\;,
\end{equation}
which we have already verified.
Therefore, two superconformal supercharges that are constructed from $\psi^-_0$ with $\gamma^+ \psi^-_0=0$ survive.

In this section, we have shown that, if $M_4$ is K\"ahler-Einstein and $F=dA$ is a harmonic anti-self-dual two-form of type (1,1) on $M_4$, it preserves two Poincar\'e supercharges when $\beta\ne 0$.
This corresponds to the kinematical supercharges.
If $\beta=0$, we additionally have two dynamical supercharges and two superconformal supercharges, adding up to six in total.
The number of surviving supercharges are the same as those of DLCQ of the AdS solution.
Note that the presence of the dynamical supercharges guarantees that the Hamiltonian $H$(the conjugate momentum to the $x^+$ coordinate) is positive definite: $\{Q, Q^{\dagger}\}=H$ for dynamical supercharges $Q$ and $Q^{\dagger}$.

\section{Specific example}\label{S:Example}

Here we present a specific example of the above analysis.
We consider the case when the four dimensional manifold $M_4$ is $S^2\times S^2$ and $y$ and $\psi$ describes a $\CP$ bundle, but warped by the $y$ coordinate.
The symmetry of the six-dimensional compact space is $SU(2)\times SU(2) \times U(1) \times \mathbb{Z}_2$ where the $U(1)$ is related to $\frac{\partial}{\partial \psi}$ and $\mathbb{Z}_2$ exchanges the two spheres.
Such a solution may be interesting since this is the symmetry of the non-relativistic limit of ABJM theory\cite{Nakayama:2009cz,Lee:2009mm}.
Let us first consider the warped $AdS_5$ solution.

\subsection{Warped $AdS_5$ solution before modification}
This solution appeared in \cite{Gauntlett:2004zh} as a specific example.
The base manifold $M_4$ is $S^2\times S^2$ of the same radius, and is a K\"ahler-Einstein manifold.
The six dimensional manifold $M_6$ has $SU(2)\times SU(2)\times U(1)$ symmetry and also a $\mathbb{Z}_2$ symmetry that switches the two $S^2$'s.
The metric is given by
\begin{equation}\label{E:WAdS5}
\begin{split}
ds_{11}^2&=e^{2\lambda(y)}\left[ds^2_{AdS_5} + ds^2_{M_6}\right]\\
ds^2_{AdS_5}&=\frac{-2 d\xp d\xm +dx_1^2+dx_2^2+dr^2}{r^2}\\
ds^2_{M_6}&=\frac 1 3 e^{-6\lambda}(1-y^2) (d\theta_1^2+\sin\theta_1^2 d\phi_1^2+d\theta_2^2+\sin\theta_2^2 d\phi_2^2)\\
&\qquad+e^{-6\lambda}\sec^2\zeta dy^2 + \frac {1}{9} \cos^2\zeta(d\psi+\hat P)^2\\
\end{split}
\end{equation}
where
\begin{equation}
\begin{split}
\hat P &= A_1+A_2\\
A_1&=-\cos\theta_1 d\phi_1\;,\qquad A_2=-\cos\theta_2 d\phi_2\\
e^{6\lambda}&=\frac{2(1-y^2)^2}{2+c y +2y^2}\\
\cos^2\zeta&=\frac{-3y^4-2c y^3-6 y^2 +1}{(1-y^2)^2}\;.
\end{split}
\end{equation}
The four-form field strength is given by
\begin{equation}\label{E:F4WAdS5}
F_4=p_1(y) \omega_1\wedge \omega_2+p_2(y) dy\wedge (d\psi+\hat P)\wedge (\omega_1+\omega_2)\;,
\end{equation}
where $\omega_i = dA_i$ and
\begin{equation}\label{E:p1p2}
p_1(y)=\frac{4y^3+3c y^2+12 y + c}{18(y^2-1)}\;,\qquad p_2(y)=\frac{y^4-6y^2-2cy-3}{9(y^2-1)^2}\;.
\end{equation}
$\theta_1$ and $\phi_1$ parametrize one $S^2$, and $\theta_2$ and $\phi_2$ the other $S^2$.
The period of $\psi$ is $2\pi$ to have a smooth geometry.
$y$ and $\psi$ combine to give a $S^2$ fibration over $S^2\times S^2$.
However, due to the $y$ dependence here and there, only $U(1)$ symmetry survives.
Also $c$ is constant, $0\le c < 4$ and $y$ runs between the two roots of the equation $\cos^2\zeta=0$. Since $\cos^2\zeta>0$ for $y=0$, one root is positive and the other negative.
It preserves 8 supercharges.

\subsection{Transformation to Schr\"odinger solution}\label{S:WSch}
Now we modify the geometry \eqref{E:WAdS5} according to section \ref{SS:DeftoSch}.
We make $\xm$ a non-trivial $U(1)$ bundle over $S^2\times S^2$ with gauge field $A=n(A_1-A_2)$, where $n$ is some integer.
The metric is given by
\begin{equation}\label{E:WSch5}
\begin{split}
ds_{11}^2&=e^{2\lambda(y)}\left[ds^2_{Sch_5} + ds^2_{M_6}\right]\\
ds^2_{Sch_5}&=-\beta y \frac{\dxp2}{r^4}+\frac{-2 d\xp (d\xm+A) +dx_1^2+dx_2^2+dr^2}{r^2}\\
ds^2_{M_6}&=\frac 1 3 e^{-6\lambda}(1-y^2) (d\theta_1^2+\sin\theta_1^2 d\phi_1^2+d\theta_2^2+\sin\theta_2^2 d\phi_2^2)\\
&\qquad+e^{-6\lambda}\sec^2\zeta dy^2 + \frac {1}{9} \cos^2\zeta(d\psi+\hat P)^2\;,
\end{split}
\end{equation}
Note that $dA$ is anti-self-dual, $dA\wedge d\hat P=0$ and $A$ does not depend on $y$.
The four-form flux is modified as follows:
\begin{equation}\label{E:F4WSch5}
\begin{split}
F_4=&p_1(y) \omega_1\wedge \omega_2+p_2(y) dy\wedge (d\psi+\hat P)\wedge (\omega_1+\omega_2)\\
&+2 n y \frac 1 {r^3} d\xp\wedge dr\wedge (\omega_1-\omega_2) -n \frac 1 {r^2} d\xp\wedge dy \wedge (\omega_1-\omega_2)\;,
\end{split}
\end{equation}
Note that the solution exists for each $c\in [0,4)$ and each integer $n$.
Given the general analysis in the previous section, the equations of motion for the four-form field and the metric are guaranteed to be satisfied.
Note that $-\beta y$, the coefficient of $\frac {1}{r^4}\dxp2$, takes both positive and negative values over the compact space.
As mentioned in section \ref{SS:DeftoSch}, this signals an instability due to the unboundedness of the Hamiltonian\cite{Hartnoll:2008rs} unless we set $\beta=0$.

Note that $dA$ is an anti-self-dual two-form of type (1,1) in $M_4$.
Hence, according to the argument in section \ref{E:genSUSY}, there are two kinematical supercharges when $\beta\neq 0$, and six supercharges when $\beta=0$. The six supercharges consist of two kinematical, two dynamical and two superconformal supercharges.
Especially, when $\beta=0$, the Hamiltonian will be bounded below due to the presence of the dynamical supercharges.

\subsection{Solution with plane wave boundary}
In the previous sections, we use the Poincar\'e coordinate system for (deformed)$AdS_5$.
In general, the $AdS_{n+2}$ metric in Poincar\'e coordinates is given by
\begin{equation}
ds^2=\frac{-2d\xp d\xm +d \vec x^2 + dr^2}{r^2}\;.
\end{equation}
where $\vec x =(x_1,\cdots, x_{n-1})$.
The boundary is $\mathbb{R}^{1,n}$.
There is another coordinate system in which the boundary approaches the plane wave metric\cite{Goldberger:2008vg,Barbon:2008bg,Blau:2009gd}.
It is given by
\begin{equation}
ds^2=\frac{-2 d{x'}^{+} d{x'}^- -\vec x'^2 d{x'}^{+2} + d\vec x'^2+ dr^2}{r^2}-d{x'}^{+2}\;.
\end{equation}
The relation between the two coordinate systems is
\begin{equation}\label{E:coordtranstopp}
\begin{split}
&x^+=\tan {x'}^+\\
&r= r' \sec {x'}^+\\
&\vec x=\vec x'\sec {x'}^+\\
&x^-={x'}^- + \frac 1 2 ({r'}^2+\vec x'^2) \tan {x'}^+\;.
\end{split}
\end{equation}
Note that $\frac{\partial}{\partial \xm}$ and $\frac{\partial}{\partial {x'}^-}$ generate the same flow in different coordinates: both are related to the number operator of the Schr\"odinger algebra.
This also suggests that not much will change even if the ${x'}^-$ direction is a line bundle over the compact space.
That is, instead of \eqref{E:WSch5}, we may consider the metric
\begin{equation}
\begin{split}
ds_{11}^2&=e^{2\lambda(y)}\left[ds^2_{Sch_5} + ds^2_{M_6}\right]\\
ds^2_{Sch_5}&=-\beta y \frac{\dxp2}{r^4}+\frac{-2 d\xp (d\xm+ A)-(x_1^2+x_2^2) \dxp2 +dx_1^2+dx_2^2+dr^2}{r^2}-\dxp2\\
&\qquad+ e^{-6\lambda}\sec^2\zeta dy^2 + \frac {1}{9} \cos^2\zeta(d\psi+\hat P)^2\;,
\end{split}
\end{equation}
This and \eqref{E:WSch5} are related by an obvious coordinate transformation.
The first term $-\beta y \frac{\dxp2}{r^4}$ may look troublesome at first, but actually $-\frac{\dxp2}{r^4}$ itself is invariant under \eqref{E:coordtranstopp}.
This form of the metric may be useful since the time direction in this coordinate system is associated to the harmonic oscillator potential
\begin{equation}
H_{osc}=H+C
\end{equation}
of the Schr\"odinger algebra.
Here $H$ generates the time translation in the Poincar\'e coordinates and $C$ is the special conformal generator.

\section{Kaluza-Klein mass spectrum}\label{S:KKmass}
The fact that the lightlike compact direction is a non-trivial bundle over the compact space has an interesting consequence on the spectrum of the Kaluza-Klein states.
We will show below that the non-relativistic particle number is bounded above by the quantum numbers of the compact space.
It seems at first a bit strange that there is such a bound.
However, we can view the system from the compact space point of view and consider the Kaluza-Klein particles charged under the momentum conjugate to the $\xm$ coordinate.
Due to the non-trivial gauge field $A$, we can think that the Kaluza-Klein particles are in a magnetic monopole background field.
Then it is well-known\cite{Wu:1976ge} that the quantum numbers of the compact space of a wave function describing a Kaluza-Klein particle is bounded below by the `electric' charge of the particle, which in this case means the $U(1)$ charge along the $\xm$ direction.
The eigenstates are expressed as monopole harmonics.
Below, we will follow the classical analysis, but in a way that can be more easily applicable to our situation.

Let us first consider the three sphere $S^3$ as a preparation.
The metric is given by
\begin{equation}\label{E:metS3}
ds^2_{S^3}=(d\psi-\cos\theta d\phi)^2 + d\theta^2 + \sin^2\theta d\phi^2\;,
\end{equation}
where $0\le \psi \le 4\pi$, $0\le \theta \le \pi$ and $0\le \phi\le 2\pi$.
The manifest symmetry is $SU(2)\times U(1)$ of $SO(4)$.
The Killing vectors are
\begin{equation}
\begin{split}
L_1&=\sin\phi \frac{\partial}{\partial \theta} + \cos\phi \left[\cot\theta \frac{\partial}{\partial \phi} +\csc\theta \frac{\partial}{\partial \psi}\right]\\
L_2&=\cos\phi \frac{\partial}{\partial \theta} - \sin\phi \left[\cot\theta \frac{\partial}{\partial \phi} +\csc\theta \frac{\partial}{\partial \psi}\right]\\
L_3&=\frac{\partial}{\partial \phi}\\
L_{\psi}&=\frac{\partial}{\partial \psi}\;.
\end{split}
\end{equation}
They satisfy $[L_i,L_j]=\sum_k \epsilon_{ijk} L_k$ and $[L_i, L_{\psi}]=0$, which comprise $SU(2)\times U(1)$ Lie algebra.
We will construct a wave function $\Phi(\psi,\theta,\phi)$ carrying definite quantum numbers of $SU(2)$ and $U(1)$.
First, let us demand
\begin{equation}
L_{\psi}\Phi=-i m_{\psi} \Phi\;.
\end{equation}
Since $\psi$ has period $4\pi$, $m_{\psi}\in \frac{\mathbb{Z}}2$.
For $SU(2)$ part, the analysis is very similar to the standard angular momentum analysis in quantum mechanics.
For the $l$ representation of $SU(2)$, let us consider the highest state $(l,m)=(l,l)$.
It will be annihilated by $L_+=L_1+i L_2$.
It is easy to see that
\begin{equation}
L_+ e^{-i m \phi} f(\theta) = 0\quad \text{for}\quad f(\theta)=\left(\frac{\sin{\frac \theta 2 }}{\cos{\frac{\theta} 2}}\right)^{m_{\psi}} \sin^l \theta\;.
\end{equation}
The wave function is then given by $\Phi(\psi,\theta,\phi)=e^{-i m_{\psi} \psi} e^{-i m \phi} f(\theta)$.
Since we want a wave function not to diverge at $\theta=0$ or $2\pi$, $|m_\psi|\le l$.
By applying the lowering operator $L_-=L_1-i L_2$ repeatedly, we obtain a wave function with definite quantum numbers $(m_{\psi},l,m)$:
\begin{equation}
\Phi_{m_{\psi},l,m}=e^{-i m_{\psi} \psi} e^{-i m \phi} (1-u^2)^{-\frac l 2} \left(\frac{1-u}{1+u}\right)^{-\frac{m_{\psi}}2}\frac{d^{l-m}}{d u^{l-m}}\left(\frac{1-u}{1+u}\right)^{m_{\psi}}(1-u^2)^{l}\;,
\end{equation}
where $u=\cos \theta$.
Since $\Phi_{m_{\psi},l,-l-k}$ has to vanish for any $k=1,2,\cdots$, $l\pm m_{\psi}$ has to be integral and positive.
In particular, $l$ can be half-integral since $m_{\psi}$ can.
The Laplacian of $S^3$ is written as
\begin{equation}
\Delta= L_1^2+L_2^2+L_3^2\;,
\end{equation}
which means the eigenvalues of the Laplacian $\Delta$ is $l(1+1)$ with $l\in \frac{\mathbb{Z}}2$.
That is, $\frac 1 4 L(L+2)$ with $L\in \mathbb{Z}$.

In sum, for a given quantum number $(l,m)$ of $SU(2)$, the possible $m_{\psi}$ range from $-l$ to $l$ with spacing $1$.
Of course, the fact that the possible values of $m_{\psi}$ are finite for a given pair of $(l,m)$ is obvious since $S^3$ has actually $SO(4)$ symmetry and for a given value of the quadratic Casimir, there are finite number of states.
However, the analysis we have done shows that the finiteness can be derived by using $SU(2)\times U(1)$ symmetry alone as well as the existence of a well-defined wave function.
For example, we would arrive at the same conclusion even though the coefficient of $(d\psi-\cos\theta d\phi)^2$ in \eqref{E:metS3} were different from $1$.

Let us turn to the case we are interested in.
The metric is given in \eqref{E:WSch5}.
There are two sets of $SU(2)$ Killing vectors $L^{(1)}_{1,2,3}$ and $L^{(2)}_{1,2,3}$ satisfying $[L^{(i)}_{a},L^{(j)}_{b}]=\sum_c \epsilon_{abc} \delta^{ij} L^{(i)}_{c}$.
Explicitly,
\begin{equation}\label{E:KVec}
\begin{split}
L^{(1)}_1&=\sin\phi_1 \frac{\partial}{\partial \theta_1} + \cos\phi_1 \left[\cot\theta_1 \frac{\partial}{\partial \phi_1} +\csc\theta_1 \left( \frac{\partial}{\partial \psi}+n \frac{\partial}{\partial \xm}\right)\right]\\
L^{(1)}_2&=\cos\phi_1 \frac{\partial}{\partial \theta_1} - \sin\phi_1 \left[\cot\theta_1 \frac{\partial}{\partial \phi_1} +\csc\theta_1 \left( \frac{\partial}{\partial \psi}+n \frac{\partial}{\partial \xm}\right)\right]\\
L^{(1)}_3&=\frac{\partial}{\partial \phi_1}\\
L^{(2)}_1&=\sin\phi_2 \frac{\partial}{\partial \theta_2} + \cos\phi_2 \left[\cot\theta_2 \frac{\partial}{\partial \phi_2} +\csc\theta_2 \left( \frac{\partial}{\partial \psi}-n \frac{\partial}{\partial \xm}\right)\right]\\
L^{(2)}_2&=\cos\phi_2 \frac{\partial}{\partial \theta_2} - \sin\phi_2 \left[\cot\theta_2 \frac{\partial}{\partial \phi_2} +\csc\theta_2 \left( \frac{\partial}{\partial \psi}-n \frac{\partial}{\partial \xm}\right)\right]\\
L^{(2)}_3&=\frac{\partial}{\partial \phi_2}\;.
\end{split}
\end{equation}
For each $S^2$, the only change from the analysis of $S^3$ is that $\frac{\partial}{\partial \psi}$ is replaced by $\frac{\partial}{\partial \psi} \pm n \frac{\partial}{\partial{\xm}}$.
Denoting the quantum numbers for $U(1)_{\psi}$ and $U(1)_{\xm}$ by $m_{\psi}$ and $N$, respectively, then we have the following constraints for given quantum numbers $(l_1,m_1;l_2,m_2)$ of $SU(2)\times SU(2)$:
\begin{equation}\label{E:mNbound}
\begin{split}
-l_1 \le m_{\psi}+n N &\le l_1\\
-l_2 \le m_{\psi}-n N &\le l_2\;.
\end{split}
\end{equation}
In particular, $N$ has to satisfy $|n N|\le l_1+ l_2$.


To see some implication of this result, let us consider the massive Klein-Gordon equation in eleven dimensions:
\begin{equation}
\frac{1}{\sqrt{-g}} \partial_M (\sqrt{-g}g^{MN}\partial_N \Phi) - m^2 \Phi=0\;.
\end{equation}
Due to the warping factor, the Laplacian becomes a little complicated.
The result can be written as
\begin{equation}\label{E:SchLap}
e^{-2\lambda}\left[-2 r^2 \frac{\partial^2\Phi}{\partial\xp\partial\xm}+r^2 \frac{\partial^2\Phi}{\partial x_1^2}+r^2 \frac{\partial^2\Phi}{\partial x_2^2}+r^5 \frac{\partial}{\partial r}\left( r^{-3} \frac{\partial \Phi}{\partial r}\right) -M^2 \Phi\right]=0\;.
\end{equation}
We put all $y$ dependence except the overall factor into a function $M^2$, which is given by
\begin{equation}\label{E:MLap}
\begin{split}
-M^2=&\beta y \frac{\partial^2 \Phi}{\partial x^{-2}}-e^{2\lambda} m^2 \Phi +\frac{e^{6\lambda}}{(1-y^2)^2} \frac{\partial}{\partial y} \left[(1-y^2)^2 \cos^2\zeta \frac{\partial \Phi}{\partial y}\right]\\
&+9 \sec^2 \zeta \frac{\partial^2\Phi}{\partial\psi^2} +\frac{3e^{6\lambda}}{1-y^2} \left[ (\Delta_1+\Delta_2)\Phi-2\frac{\partial^2\Phi}{\partial \psi^2} - 2n^2 \frac{\partial^2\Phi}{\partial x^{-2}}\right]\;.
\end{split}
\end{equation}
$\Delta_1$ and $\Delta_2$ are the Casimir operators of the two $SU(2)$ isometry groups, which are given by $\Delta_{i}=(L^{(i)}_1)^2+(L^{(i)}_2)^2+(L^{(i)}_3)^2$ using \eqref{E:KVec}.
For a wave function with definite quantum numbers of $SU(2)\times SU(2)\times U(1)_{\psi}$ and definite particle number, this equation becomes an ordinary second order differential equation in $y$.
Note that the last term in \eqref{E:MLap} looks problematic since, by increasing the momenta along the $\psi$ and $\xm$ directions, this part can be negative and large in absolute value.
However, this cannot happen since the quantum numbers $m_{\psi}$ and $N$ are bounded.
That is, from \eqref{E:mNbound}, we have
\begin{equation}
l_1(l_1+1)+l_2(l_2+1)\ge l_1^2 + l_2^2 \ge (m_{\psi}+n N)^2 + (m_{\psi}-n N)^2 = 2m_{\psi}^2 + 2 n^2 N^2\;.
\end{equation}
It implies that the operator
\begin{equation}\label{E:Oexp}
\mathcal{O}=\Delta_1+\Delta_2-2\frac{\partial^2}{\partial \psi^2} - 2n^2 \frac{\partial^2}{\partial x^{-2}}
\end{equation}
cannot have positive eigenvalues.
Therefore, the last three terms in \eqref{E:MLap}(multiplied by $e^{-2\lambda}$) gives positive contribution to the mass parameter $M^2$.
That is, when $\beta$ vanishes, the Kaluza-Klein mode does not suffer an instability due to the violation of the Breitenlohner-Freedman bound.

If we solve \eqref{E:MLap} and get the spectrum of the mass parameter $M$, the scaling dimensions and the correlation functions can be computed\cite{Son:2008ye,Balasubramanian:2008dm}.
Let $\nu=\sqrt{M^2+4}$. The scaling dimension $\Delta$ of the corresponding operator in the field theory is given by $\Delta=2+\nu$ and the two point correlation function of two such operators is given by
\begin{equation}
\left< \mathcal{O}_1 (x,t) \mathcal{O}_2 (0,0)\right>\sim \delta_{\Delta_1 \Delta_2} \theta(t) \frac{1}{t^{\Delta_1}} e^{-\frac{i N x^2}{2t}}\;,
\end{equation}
where $\Delta_i$ are the scaling dimensions of $\mathcal{O}_i$. $\Delta=2-\nu$ is possible if $0<\nu<1$\cite{Son:2008ye,Klebanov:1999tb}.

\section{Solution with no supersymmetry}
In the absence of supersymmetry, there may be many solutions with the symmetries we want.
The solution given here can be thought of as a deformation of the non-supersymmetric $AdS_5\times \CPPP$ solution in \cite{Pope:1988xj}.
As such, the solution here does not preserve any supersymmetry.
We simply state the solution since it is straightforward to check that the solution satisfies the equations of motion.
We take the lightlike direction $\xm$ to be a non-trivial $U(1)$ bundle over the compact direction.
That is, an invariant combination is $d\xm+ n A$ where $A$ is a gauge potential given below.
For each integer $n$ there is a solution.
The metric is given by
\begin{equation}
\begin{split}
ds^2=&-10 n^2 \frac{\dxp2}{r^4} + \frac{-2d\xp (d\xm+ n A)+ dx_1^2+ dx_2^2 + dr^2}{r^2} +\frac 1 2 ds_{N_6}^2\\
ds^2_{N_6}=& \frac{d\alpha^2}{f(\alpha)}+f(\alpha)\sin^2\frac{\alpha} 2 \cos^2\frac {\alpha}2(d\chi + \cos \theta_1 d\phi_1 - \cos \theta_2 d\phi_2)^2\\
&+ \cos^2 \frac{\alpha} 2 (d\theta_1^2 + \sin^2 \theta_1^2 d\phi_1^2)+\sin^2 \frac{\alpha} 2 (d\theta_2^2 + \sin^2 \theta_2^2 d\phi_2^2)\\
A=&\cos\theta_1 d\phi_1 + \cos\theta_2 d\phi_2+\cos\alpha (d\chi + \cos\theta_1 d\phi_1-\cos\theta_2 d\phi_2)\\
f(\alpha)&=1- \frac k {\sin^4\alpha}\;.
\end{split}
\end{equation}
$k$ is some constant and $\theta_i \in [0,\pi]$, $\phi_i \in [0,2\pi]$ and $\chi \in [0,4\pi]$.
The four-form field strength $F_4$ is given by
\begin{equation}
F_4= \frac{\sqrt 2}{16} \omega_2\wedge \omega_2 + \frac n {\sqrt 2} \frac 1 {r^3} d\xp\wedge dr \wedge \omega_2 +12n \frac 1{r^5} d\xp\wedge dx_1 \wedge dx_2\wedge dr\;,
\end{equation}
where $\omega_2=dA$ is proportional to the K\"ahler form.
$N_6$ is a six-dimensional compact manifold.
It is a variant of $\CPPP$: When $k=0$, $N_6$ becomes $\CPPP$.
$k$ is fixed once we require the manifold $N_6$ to be smooth.
If $k=0$, $N_6$ is smooth since it is $\CPPP$, in which case the global symmetry is $SU(4)$.
To get reduced symmetry, we want to take non-zero $k$.
For non-zero $k$, since $f(\alpha)$ is supposed to be positive, $\alpha$ runs between the two roots of $\sin^4 \alpha=k$.
Calling the roots $\pm \alpha_0$, near $\alpha_0$, the metric becomes
\begin{equation}
ds^2=\tan\alpha_0\left[du^2 + \cos^2\alpha_0 u^2 (d\chi+\cos\theta_1 d\phi_1 - \cos\theta_2 d\phi_2)^2\right]+\cdots\;,
\end{equation}
where $\alpha=\alpha_0+u^2$.
Since $\chi$ has period $4\pi$, $\cos\alpha_0=\frac 1 2$ to have a smooth geometry.
Therefore, $k=\frac 9 {16}$ and $\alpha\in [\frac {2\pi} 3, \frac {4\pi}3]$.
In this case, the surviving global symmetry is $SU(2)\times SU(2)\times U(1)\times \mathbb{Z}_2$.
Note that this construction can be easily generalized to the case when the compact six-dimensional manifold is K\"ahler-Einstein.

\section{Discussion}
In this section, we present one possible way of achieving Galilean symmetry from Poincar\'e symmetry, which gives a reason why we consider the case where the lightlike direction is a $U(1)$ bundle over the compact space.

Let us consider a non-relativistic limit of some geometry in general, which has $(1+d)$-dimensional Poincar\'e symmetry.
It need not have scale invariance.
Then, it has translational symmetries, whose associated Killing vectors are $\frac{\partial}{\partial t}$, $\frac{\partial}{\partial x_i}$.
Additionally, there are rotational symmetries. Killing vectors for spatial rotations are $x_i \frac{\partial}{\partial x_j}-x_j \frac{\partial}{\partial x_i}$ and those for Lorentz boosts are $t\frac{\partial}{\partial x_i}+x_i \frac{\partial}{\partial t}$.
Suppose also that there is a $U(1)$ isometry and $\frac{\partial}{\partial \phi}$ is the corresponding Killing vector.

Now, one way to send this geometry to a non-relativistic limit is to change the coordinates so that, instead of $\{t,x_i,\phi\}$, we use $\{t', x'_i, \phi' \}$ as coordinates where
\begin{equation}\label{E:coordtrans}
\begin{split}
t'&=\epsilon^2 t\\
x'_i&=\epsilon x_i\\
\phi'&=\phi+ t\;.
\end{split}
\end{equation}
$\epsilon$ is some constant.
Correspondingly, the Killing vectors are expressed in this new coordinate system as
\begin{equation}
\begin{split}
\frac{\partial}{\partial t}&=\epsilon^2 \frac{\partial}{\partial t'}+\frac{\partial}{\partial \phi'}\\
\frac{\partial}{\partial x_i}&=\epsilon \frac{\partial}{\partial x'_i}\\
\frac{\partial}{\partial \phi}&=\frac{\partial}{\partial \phi'}\;.
\end{split}
\end{equation}
Translational symmetries along $t,x_i,\phi$ directions and space rotational symmetries are expressed in the same way as in the original coordinate system.
But the Lorentz boost symmetries in the new coordinate system are
\begin{equation}\label{E:LorentzBoost}
V_i=\epsilon\left( t\frac{\partial}{\partial x_i}+x_i \frac{\partial}{\partial t} \right)=t' \frac{\partial}{\partial x'_i} + x'_i \frac{\partial}{\partial \phi'} + \epsilon^2 x'_i \frac{\partial}{\partial t'}\;.
\end{equation}
Since $\epsilon$ is just some constant, it can be freely multiplied to the generators as above.
Given this form, take the limit $\epsilon\rightarrow 0$.
Then the resulting generators are
\begin{equation}
V_i|_{\epsilon\rightarrow 0}=t' \frac{\partial}{\partial x'_i} + x'_i \frac{\partial}{\partial \phi'}\;,
\end{equation}
which are the Galilean boost symmetry generators.

This seems that we should be able to obtain the Galilean boost symmetry generators when we take a certain limit of a geometry with Poincar\'e symmetry.
But, of course, we have glossed over an important requirement that the geometry should behave well when we take the coordinate transformation \eqref{E:coordtrans} and take the limit $\epsilon\rightarrow 0$.

Before examining an example, let us comment on the coordinate transformation \eqref{E:coordtrans}.
First, $\phi$ in \eqref{E:coordtrans} has been shifted by $t$, but in general it can be replaced by $k t$ with some constant $k$.
But we do not take a special limit for $k$.
Note that $t$ and $x^i$ are rescaled, but $\phi$ is not.
The reason is that we have in mind the case where $\phi$ is compact, in which case rescaling is not possible.
Also, shifting $\phi$ coordinate by $t$ does not spoil the periodicity of $\phi$.

The example considered in the following is the near M5-brane geometry of the LLM geometry with M2-branes polarized into a small number of M5-branes\cite{Bena:2004jw,Lin:2004nb}.
The metric takes the form
\begin{equation}
ds_{11}^2=(\pi N_5)^{-\frac 1 3} r (-dt^2+d{x}_1^2+d{x}_2^2 + ds_{S^3}^2) + (\pi N_5)^{\frac 2 3}\frac 1 {r^2} \left[dr^2 + r^2 (d\theta^2+\sin^2\theta ds_{\tilde{S}^3}^2)\right]\;.
\end{equation}
$N_5$ is the number of polarized $N_5$ branes.
M5-branes are wrapping $t,x_1,x_2$ and $S^3$.
As $r$ becomes large, the geometry becomes $AdS_7\times S^4$.
Although we do not have an exact field theory dual for multiple M2-branes in flat space with mass deformation, we expect that this geometry can be thought of as a vacuum in which matter fields have non-zero expectation values along the first four coordinates in $\mathbb{R}^8$\cite{Bena:2000zb,Bena:2004jw,Lin:2004nb}.
As such, the geometry will not have a relativistic or non-relativistic scaling symmetry and we do not expect it to have a well-defined limit of the previous discussion.
However, we can just see where the limit sends this geometry to.

Writing the metric for the three sphere $S^3$ as
\begin{equation}
ds_{S^3}^2=(d\phi+ A)^2+ ds_{\CP}^2\;,
\end{equation}
we apply the coordinate transformation \eqref{E:coordtrans}.
Then we have
\begin{equation}
ds_{11}^2=(\pi N_5)^{-\frac 1 3} \frac{r}{\epsilon^2} (-(d\phi'+A) d t' + d{x_1}'^2+d{x_2}'^2 + \epsilon^2 ds_{S^3}^2) + (\pi N_5)^{\frac 2 3}\frac 1 {r^2} \left[dr^2 + r^2 (d\theta^2+\sin^2\theta d\tilde{s_{S^3}}^2)\right]\;.
\end{equation}
If we further change the coordinate $r$ to $r'=\frac r {\sqrt\epsilon}$,
\begin{equation}
ds_{11}^2=(\pi N_5)^{-\frac 1 3} r' (-(d\phi'+A) d t' + d{x_1}'^2+d{x_2}'^2 + \epsilon^2 ds_{S^3}^2) + (\pi N_5)^{\frac 2 3}\frac 1 {r'^2} \left[dr'^2 + r'^2 (d\theta^2+\sin^2\theta d\tilde{s_{S^3}}^2)\right]\;.
\end{equation}
Note that the $U(1)$ bundle associated with the $\phi'$ coordinate becomes lightlike when $\epsilon\rightarrow 0$.
However, as expected, the limit $\epsilon\rightarrow 0$ does not exist since the metric $d\Omega_3^2$ for $S^3$ shrinks to zero.
But, for small $\epsilon$, this geometry can be thought of as a compactification and the Lorentz boost generators \eqref{E:LorentzBoost} becomes approximately the Galilean boost generators for small $\epsilon$.

In the case of the mass deformed ABJM theory, the analogue of the LLM geometry is not known.
Especially, we lack a geometry dual to the most symmetric vacuum of the field theory, in which the matter fields have vanishing expectation values.
If we have such a solution, we will be able to check whether we can apply the procedure mentioned above to that solution.

\section*{Acknowledgments}
We would like to thank Jaewon Song and Masahiko Yamazaki for their participation in an earlier stage of this project and for
discussion.
We are grateful to Yu Nakayama for helpful discussion. 
This work is supported in part by DOE grant DE-FG03-92-ER40701. The work of H.O. is
also supported in part by a Grant-in-Aid for Scientific Research (C) 20540256 from the Japan
Society for the Promotion of Science, by the World Premier International Research Center
Initiative of MEXT of Japan, and by the Kavli Foundation.
C.P. is supported in part by Samsung Scholarship.

\appendix

\section*{Appendix}

\section{Notation}

We mostly follow the notation of \cite{Becker:2007zj}.
In the supergravity approximation of M theory, the Lagrangian is given by
\begin{equation}
\mathcal{L}=\frac{1}{2\kappa_{11}^2}\left[ \int d^{11} x \sqrt{-g} \left( R- \frac 1 2 |F_4|^2\right) - \frac 1 6 \int A_3 \wedge F_4\wedge F_4 \right]\;.
\end{equation}
The quantity $|F_p|^2$ is defined by
\begin{equation}
|F_p|^2 = \frac{1}{p!} g^{M_1 N_1} \cdots g^{M_p N_p} F_{M_1\cdots M_p}F_{N_1\cdots N_p}\;.
\end{equation}
Indices $M,N,\cdots$ run from 1 to 11 and denote coordinate indices.
The metric is mostly positive.
The vielbein indices are denoted by $A,B,\cdots$.
The equation of motion for $A_3$ is
\begin{equation}\label{E:EOMF4}
\begin{split}
d F_4&=0\\
d* F_4 + \frac 1 2 F_4 \wedge F_4&=0\;.
\end{split}
\end{equation}
The equation of motion for the metric $g_{MN}$ is
\begin{equation}\label{E:EinsteinEq}
G_{MN} = \kappa_{11}^2 T_{MN}\;,
\end{equation}
where $G_{MN}$ is the Einstein tensor and
\begin{equation}
T_{MN}=-\frac{2}{\sqrt{-g}} \frac{\delta S_{A_3}}{\delta g^{MN}}\;,
\end{equation}
where $S_{A_3}$ denotes the part of the action excluding the Ricci scalar term.
Explicitly,
\begin{equation}\label{E:TMNF}
T_{MN}=-\frac 1 {4\kappa_{11}^2} \left(\frac 1 {4!} g_{MN} F^{M_1\cdots M_4} F_{M_1\cdots M_4} - \frac 2 {3!} F_{M M_1 M_2 M_3} F_{N}^{\;\;\; M_1 M_2 M_3}\right)\;.
\end{equation}
In terms of the gamma matrices,
\begin{equation}\label{E:TMN}
T_{MN}=-\frac 1 {4\kappa_{11}^2} \frac 1 {32} \Tr(\Gamma_M \bF^{(4)} \Gamma_N \bF^{(4)})\;.
\end{equation}
where $\bF^{(4)}=\frac 1 {4!} F_{MNPQ} \Gamma^{MNPQ}$.
The eleven gamma matrices satisfy the relation
\begin{equation}
\{ \Gamma^A, \Gamma^B\}=2\eta^{AB}\;,
\end{equation}
where $\eta^{AB}=\mathrm{diag}(-1,1,\cdots,1)$.

%
%
%
%

\end{document}